\documentclass[10pt,journal,compsoc,letterpaper]{IEEEtran}
\usepackage[utf8x]{inputenc}

\usepackage{cite}
\usepackage{tabularx,booktabs}
\usepackage{multirow}
\usepackage{verbatim}

\usepackage[cmex10]{amsmath}
\usepackage{amssymb}

\usepackage{xspace}

\usepackage[inline]{enumitem}

\usepackage{hyperref}

\usepackage{siunitx}
\usepackage{upquote}
\usepackage{listings}

\usepackage{graphicx} 
\ifCLASSOPTIONcompsoc
    \usepackage[caption=false, font=normalsize, labelfont=sf, textfont=sf]{subfig}
\else
\usepackage[caption=false, font=footnotesize]{subfig}
\fi

\usepackage[framemethod=TikZ]{mdframed}
\mdfdefinestyle{style1}{
innerleftmargin=0.5cm,innerrightmargin=0.4cm,
innertopmargin=0.3cm ,innerbottommargin=0.3cm,
roundcorner=10pt,linewidth=1.0pt,
footnoteinside=false}

\newcolumntype{Y}{>{\centering\arraybackslash}X}
\newcolumntype{Z}{>{\hsize=.5\hsize}X}

\newcommand{\mr}{\mathrm}

\newcommand{\nfaults}{\textit{faults}\xspace}
\newcommand{\approach}{\textit{approach}\xspace}
\newcommand{\experience}{\textit{experience}\xspace}

\newcommand{\expect}{\ensuremath{\mathbb{E}}}

\newcommand{\model}{\ensuremath{\mathcal{M}}}

\newcommand{\mypara}[1]{\textbf{#1.}}

\begin{document}




\title{A Method to Assess and Argue for \\Practical Significance in Software Engineering}

%
%
%

\author{Richard Torkar,
        Carlo A.~Furia,
        Robert Feldt,
        Francisco Gomes de Oliveira Neto,\\
        Lucas Gren,
        Per Lenberg,
        and Neil A.~Ernst
\thanks{R.~Torkar, R.~Feldt, L.~Gren, P.~Lenberg, and F.~Gomes de Oliveira Neto are with the Department of Computer Science and Engineering, Chalmers and University of Gothenburg, Sweden. e-mail: torkarr@chalmers.se.}
\thanks{R.~Torkar is also with the Stellenbosch Institute for Advanced Study (STIAS), Wallenberg Research Centre at Stellenbosch University, Stellenbosch, South Africa.}%
\thanks{C.~A.~Furia is with the Software Institute of USI Universit\`a della
Svizzera italiana,
Via Giuseppe Buffi 13,
I-6904 Lugano,
Switzerland.}
\thanks{L.~Gren is also with Blekinge Institute of Technology, Karlskrona, Sweden.}%
\thanks{N.~A.~Ernst is with the Department of Computer Science, University of Victoria, 3800 Finnerty Rd, Victoria, BC V8P 5C2, Canada}}

\markboth{IEEE Transactions on Software Engineering,~Vol.~XX, No.~X, Month~YYYY}{}%


\IEEEtitleabstractindextext{%
\begin{abstract}
A key goal of empirical research in software engineering is to assess practical significance, which answers whether the observed effects of some compared treatments show a relevant difference in practice in realistic scenarios. Even though plenty of standard techniques exist to assess statistical significance, connecting it to practical significance is not straightforward or routinely done; indeed, only a few empirical studies in software engineering assess practical significance in a principled and systematic way.

In this paper, we argue that Bayesian data analysis provides suitable tools to assess practical significance rigorously. We demonstrate our claims in a case study comparing different test techniques. The case study's data was previously analyzed (Afzal et al., 2015) using standard techniques focusing on statistical significance. Here, we build a multilevel model of the same data, which we fit and validate using Bayesian techniques. Our method is to apply cumulative prospect theory on top of the statistical model to quantitatively connect our statistical analysis output to a practically meaningful context. This is then the basis both for assessing and arguing for practical significance. 

Our study demonstrates that Bayesian analysis provides a technically rigorous yet practical framework for empirical software engineering. A substantial side effect is that any uncertainty in the underlying data will be propagated through the statistical model, and its effects on practical significance are made clear.

Thus, in combination with cumulative prospect theory, Bayesian analysis supports seamlessly assessing practical significance in an empirical software engineering context, thus potentially clarifying and extending the relevance of research for practitioners. 
\end{abstract}

\begin{IEEEkeywords}
practical significance, statistical significance, Bayesian analysis, empirical software engineering
\end{IEEEkeywords}
}

\maketitle


\IEEEraisesectionheading{\section{Introduction}\label{sec:intro}}
\IEEEPARstart{A}{main} goal of research in empirical software engineering (ESE) 
is assessing practical significance: what is the impact of the research findings
in realistic scenarios?
To this end, statistical analysis has been used extensively in ESE for decades.
Nonetheless, the bulk of research has focused on
defining and implementing guidelines for 
experimental design (from case studies~\cite{Runeson2008casestudies} 
to grounded theory~\cite{Stol2016groundedtheory} and experiments~\cite{Wohlin2012book})
and statistical analysis (from statistical testing~\cite{Arcuri2011hitchhiker}
to Bayesian modeling~\cite{Furia2018BDA}).
In contrast, practical significance is rarely discussed explicitly or quantitatively~\cite{deOliveiraNeto2019Statistics}.

The most common approach to assessing the significance of findings 
is built on top of statistical significance, which is extended in a quantitative way.
A common example are effect size measures (such as Cohen's $d$, or the size of coefficients in a regression model):
if the effect size of a technique $A$ is markedly bigger than the one of another technique $B$,
this is taken as an indication that $A$ performs better than $B$ in practice.
This common approach overlooks the issue that 
assessing practical significance on statistical measures such as effect sizes
makes it hard to ensure that the statistics accurately reflect expert knowledge.
In particular, practitioners (who are the experts) may not be familiar with the nuances of
the various statistical techniques and how they are used. 

Furthermore, showing statistically that one technique performs better than another one does not automatically mean that this makes a difference in practice.
Using effect sizes frames practical significance as a general property~\cite{Tantithamthavorn2018icse-seip};
however, it is much more likely to be a context-dependent property, as whether a technique 
will be better than another in practice depends on the context where those techniques will be
deployed.
Therefore, 
a quantitative assessment of practical significance should be expressible 
in terms of
(or, at least, clearly connected to)
measures in the application domain that are used by the domain experts.
For example, return on investment, time, and personnel costs are all measures that
are appropriate for an evaluation of economic impact.

Grounding practical significance in domain-specific metrics
also supports a clear communication of the expected impact
of the solutions that have been empirically assessed;
in other words, how each solution would help practitioners~\cite{Storey2017_esem},
and how they could choose the one most appropriate for their needs~\cite{Petersen2014_serp,Siegmund2015_icse}.
Establishing such a clear communication would 
ultimately enhance research's long-term impact~\cite{Petersen2014_serp,Engstrom2017_serptest}.

In this paper, we demonstrate how 
a combination of Bayesian analysis~\cite{mcelreath15statrethink,gelman2013bayesian,Furia2018BDA} and cumulative prospect theory~\cite{Tversky1992cpt,KahnemanT79prosp}
provides a serviceable means of assessing practical significance in ESE\@.
Using data from a case study of testing practices~\cite{AfzalGITAB2015et} 
(comparing exploratory testing to testing based on predefined test cases),
we illustrate how one can formulate and assess different measures of practical significance,
all expressed in terms of metrics that make sense in the application domain
(such as the hourly cost and seniority of programmers).

We use Bayesian statistics to design and fit a model of the empirical data.
The model is quantitative, incorporates expert knowledge in the domain metrics of interest, 
and can be used to perform predictions.
We then use cumulative prospect theory to connect probabilities in the Bayesian model to utility metrics of possible outcomes. The connection is actionable, in that it supports decision making based on the applicability, risks, and costs of different scenarios. If one would take this one step further, these costs per scenario could greatly reduce the need for human subjects for validating research results as cumulative prospect theory will provide us with strong indications of what decision a human will take.

\subsection{Cumulative Prospect Theory}

Cumulative prospect theory (CPT) is a framework developed in behavioral economics
to model decisions under risk and uncertainty~\cite{Tversky1992cpt} and has been applied for practical decision making, e.g. in medicine~\cite{bayoumi2000decision}. 
CPT models several aspects of human decision making, 
such as sensitivity to how options are framed, non-linear sensitivity to risk, and loss aversion;
based on these factors, a CPT model defines the utility of a certain decision's outcome.

Software engineering practitioners are faced with decision making under uncertainty, so CPT is also a useful framework in this domain.
As a simple example, consider a manager who is organizing a code review.
Two options are available to them: approach $A$, which guarantees a value of \$940; and approach $B$, which gives a value of \$1000 with 95\% probability, and no value (\$0) with 5\% probability.
CPT indicates that most managers will choose $A$ given its certainty, even though the expected utility of $B$ is slightly better (as $0.95 \times 1000 = \$950 > \$940$). This behavior is a manifestation of risk aversion, which may also depend on how the problem is framed.

CPT can provide a suitable model of how decisions are made in software engineering practice as well.
As we show in this paper, 
empirical data can be used to fit probabilistic models of different outcomes;
the probabilities correspond to risks in a CPT model.
The latter also acts as a sort of ``high-level interface'' for the 
practitioners and decision makers, who do not have to understand the statistical model
but can reason in the more familiar terms of risks and utility values of each possible outcome.
Domain expertise remains crucial to build a suitable CPT model:
in the previous example of code review approaches $A$ and $B$,
domain experts would estimate the profits and costs associated with each option.

\begin{figure*}[!hbt]
    \centering
    \includegraphics[width= \textwidth]{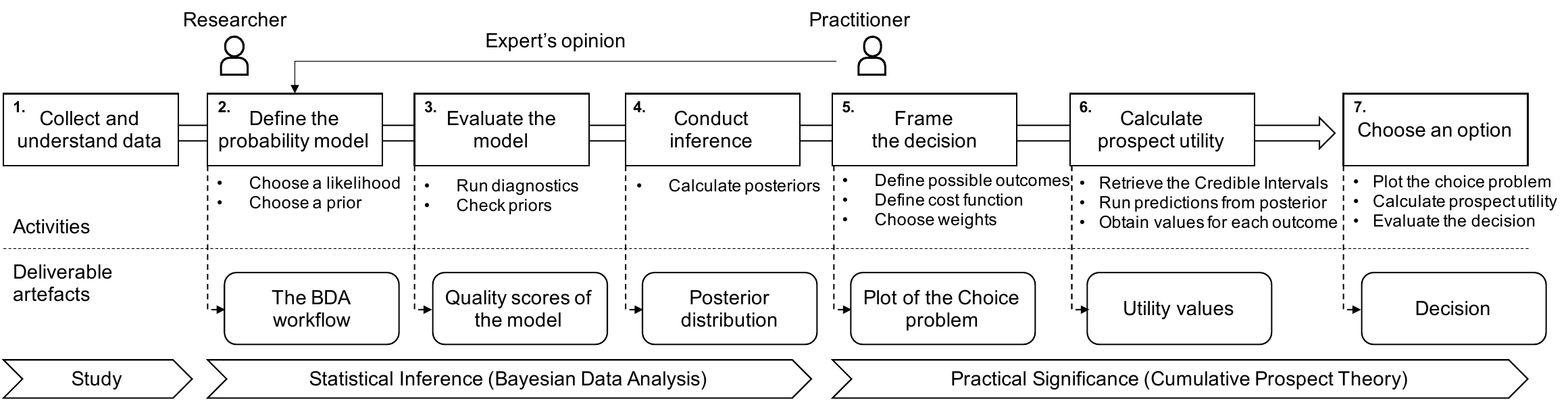}
    \caption{Assessing practical significance using a combination of Bayesian analysis and cumulative prospect theory.}
    \label{fig:our_approach}
\end{figure*}

\subsection{Proposed Solution}
We combine cumulative prospect theory and Bayesian statistics to assess practical significance
of ESE data.
Figure~\ref{fig:our_approach} gives an overview of our approach.
We use CPT to model the possible decisions, how they are framed, and the risks associated with them.
Then, we use Bayesian statistical analysis to infer the probabilities that quantify the risk of each decision.
While in principle any statistical approach could work,
Bayesian models are better suited
because they provide 
a detailed \emph{posterior}
distribution of the possible outcomes (instead of just point or interval estimates),
which can be seamlessly combined with CPT models.
In addition, Bayesian models 
are easy to interpret~\cite{Furia2018BDA}
and naturally incorporate expert knowledge and assumptions through the use of \emph{priors}.

Our approach is applicable to many ESE topics and subject areas, as long as the investigation is suitable for a quantitative analysis based on statistical methods. In this paper, we do not investigate the connection between our proposed framework and qualitative analysis in empirical software engineering, as both approaches are complementary. In other words, we assume that there are strategies to identify and measure values that can be translated into decisions framed as weighting functions. 

\begin{figure*}
    \centering
    \includegraphics[width= \textwidth]{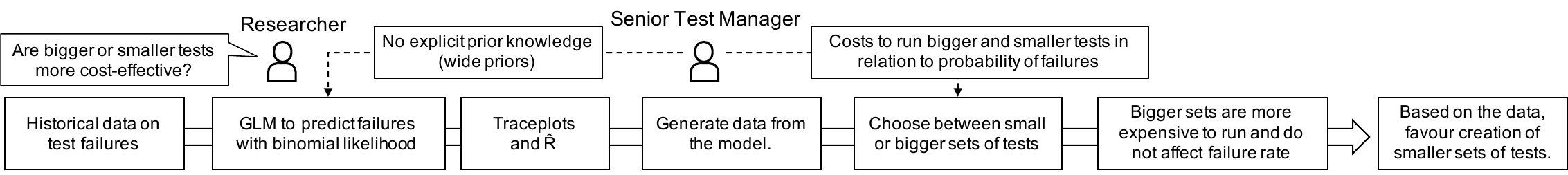}
    \caption{An illustration of our approach using BDA and CPT applied to the decision between bigger or smaller test suites for cost-effective testing. The GLM enables the prediction of failures based on the test suite sizes via a Binomial likelihood. The expert (senior test manager) provides the cost model of running a test suite (e.g., including costs for both size and fault fixing) that is then framed as choices between test suite sizes with corresponding utility values. Practitioners then choose the outcome with better prospect utility.}
    \label{fig:approach_example}
\end{figure*}

As an illustration (Figure~\ref{fig:approach_example}), consider that we investigate whether to favour the execution of a bigger, costlier, but more thorough test suite versus smaller, cheaper, and relatively superficial test suites. Using data collected from mining software repositories and an expert's opinion, we can create a generalized linear model (GLM) to analyse how the different sizes of test suites affect executions costs, as well as the number of failures that are revealed and must be fixed. Ultimately, the model provides input to calculate the utility of both types of test suites in connection to different probable scenarios, such as the test execution costs based on the unlikely situation of revealing too many failures.

We evaluate the feasibility of this approach in three steps.
First, we reanalyze a previously published empirical study~\cite{AfzalGITAB2015et}
using Bayesian techniques.
Second, we combine the Bayesian model with CPT under plausible practical application scenarios. 
Third, we ask practitioners to compare the information provided
by the original study~\cite{AfzalGITAB2015et}, to that derived by our
combination of Bayesian statistics and CPT, and to 
indicate which better supports their decisions.
Overall, we demonstrate how to build a rigorous model of practical significance, 
and the advantages of reporting significance results in a way that is grounded in concrete decision-making scenarios---in contrast to the traditional approach 
that presents general statistics in a more abstract form.

In summary, the contributions of this paper are:

\begin{itemize}
    \item An approach that connects statistical inference and practical significance to frame empirical findings in a way that supports practitioners in making decisions.
    \item A case study of applying cumulative prospect theory in ESE.
    \item A reanalyses of previously published data~\cite{AfzalGITAB2015et},
      which revisits the original findings and extends them to a context of practical significance.
    Our analysis is reproducible and available online~\footnote{\url{https://github.com/torkar/docker-b3}}.
\end{itemize}

This paper is structured as follows. Section~\ref{sec:background} presents the background for both Bayesian analysis and CPT\@,
so as to introduce readers to the essential terminology
and steps of both techniques.
We present our analysis in Section~\ref{sec:experiment_bda}, followed by the application of CPT in Section~\ref{sec:experiment_cpt}. The results from the validation of our approach with practitioners are presented in Section~\ref{sec:validation}.
In Section~\ref{sec:discussion} we discuss the implications of our work, while in Sections~\ref{sec:threats} and~\ref{sec:conclusion} we present threats to validity and conclude the paper.

\section{Background and Related Work}
\label{sec:background}

In order to present the background of our research, we introduce the essential terminology and techniques of Bayesian statistical analysis
and CPT\@. 
We also discuss related work in ESE literature that also targets the practical impact of research findings.

\subsection{Bayesian Analysis}
\label{sec:background:bda}

Data analysis relies on statistical analysis to infer properties of a population that follows an underlying probability distribution. 
Since the actual underlying distribution is often unknown, statistical inference
estimates probabilities by generalizing the frequencies observed in finite \emph{samples} of the population.

There exist different families of probability distributions, such
as uniform, normal, binomial, Poisson, beta, etc.
Each distribution in a certain family is characterized by the values of one or more parameters 
$\theta_1, \ldots, \theta_n$, 
which fix the distribution's shape.
For instance, a normal (also called Gaussian) distribution has two parameters: mean $\mu$ and standard deviation $\sigma$. 
The key goal of statistical analysis is to estimate a distribution parameter $\theta$
from \emph{sampled data} $D$ that was drawn from the population.
In Bayesian statistics, 
this is expressed as estimating the probability $P(\theta \mid D)$ of the parameters given the data,
which obeys Bayes' theorem:

\begin{equation}
\label{eq:bayes_theorem}
P(\theta \mid D) = \frac{P(D \mid \theta) \times P(\theta)}{P(D)}\,\text{,\quad where:}
\end{equation}

\begin{itemize}
    \item $P(\theta \mid D)$ is the \textit{posterior distribution} and is what we want to estimate. 
    
    \item $P(D \mid \theta)$ is the \textit{likelihood} that the data was drawn from a distribution with parameters $\theta$. 
    
    \item $P(\theta)$ is the \textit{prior}, 
    which encodes prior knowledge by constraining plausible values for the parameters $\theta$ independent of the data.
    \item $P(D)$ is simply a normalizing constant.
\end{itemize}

As a concrete example, imagine that we are investigating the number of tests failing during a build of a certain project.\footnote{For the sake of simplicity, we assume that a test either passes or fails (no flaky behavior or timeouts are possible).} We would like to infer the failure rate, i.e., the probability $p_i$ of a test failing in an arbitrary build. 
The binomial distribution family represents sequences of $n$ events each with a fixed probability of success $p$.
In our example, the $n$ events are the tests that are executed in each build,
and $p_i$ is the probability of failure for test $i$.
Thus, draws from the distribution Binomial$(n, p)$ 
capture the likelihood of observing a certain number of tests failing given parameters $n$ and $p$.

The case study in Section~\ref{sec:experiment_bda} demonstrates these concepts in greater detail. 
For additional technical details about how Bayesian analysis is applied, we refer to \cite{Furia2018BDA} and \cite{mcelreath15statrethink}. The key features of a Bayesian approach, which makes it best suited to analyze practical significance are:

\begin{enumerate}
    \item It can incorporate prior knowledge.
    \item It produces a posterior predictive distribution (PPD).
    \begin{enumerate}
        \item The PPD is conditioned on the observed data.
        \item The conditioning allows us to update our beliefs of the unknowns.
        \item The variation indicates any remaining uncertainty in our beliefs about the unknowns.
        \item It produces estimates of the standard deviation of the marginal posterior distributions.
    \end{enumerate}
    
    \item It produces Bayesian uncertainty intervals, 
    which quantitatively measure degree of belief after seeing the data.
\end{enumerate}

\subsection{Cumulative Prospect Theory}
\label{sec:background:cpt}

A fitted Bayesian model can be used to \emph{simulate} new scenarios that generalize
those in the observed data.
Each scenario is associated with different outcomes, costs, and potential benefits.
In the example of testing outlined in the previous section,
different testing practices lead to different failure rates and
have different application costs (for example because they require more developers);
on the other hand, failing tests have to be fixed, which introduces additional costs.
We use cumulative prospect theory (CPT)~\cite{Tversky1992cpt}
to support a decision-making process, which is based on scenarios that are modeled statistically.

CPT is a widely used decision-making modeling framework, which accounts for 
experimentally demonstrated features of human behavior when making a decision:

\begin{itemize}

\item\textit{Loss aversion} is the preference for avoiding losses over gaining advantages---even when the latter would outweigh the former.

\item \textit{Framing effects} refer to the widespread psychological phenomenon that the same data 
may lead to very different choices according to \emph{how} (in which scenario) the data is presented.

\item \textit{Nonlinear preferences:} 
  humans associate risk to the 0--100\% probability scale in a way that is not uniform.
  For example, the difference between a 99\% and 100\% risk is considered more significant
  than the difference between a 10\% and 11\% risk.

\item People tend to be \emph{risk averse}, except in specific scenarios where they are 
more likely to actively \emph{seek risks}: 
when there is a small probability of winning a large prize
and when choosing between a sure loss and a substantial probability of an even larger loss.

\item \textit{Source dependence} refers to the influence of domain expertise on preferences.
Decision makers usually feel more confident within their area of expertise even when they have limited and noisy data---often even more confident than when they have detailed data about an area they are not familiar with~\cite{Heath1991preferences}.
\end{itemize}

Concretely, CPT represents
a decision using a function $\nu: X\rightarrow \mathbb{R}$
that maps each possible outcome $x \in X$ of the decision to its value $\nu(x)$.
A positive value is a gain, and a negative one is a loss.
We associate each outcome $x$ with a weight that reflects the outcome's probability
corrected to account for how it is subjectively perceived (according to the phenomena listed above).
More precisely, CPT provides standard weighting functions that can be applied to any probability distributions of outcomes to derive the ``subjective'' probability associated with any outcome.
The subjective expected utility value $\expect_U(X)$ of the decision 
is the weighted average of the value of
all possible outcomes---each weighted by its subjective probability, 
which in turn is based on the outcome's ``objective'' probability (from a statistical model).

Continuing the example of the failing tests,
we imagine a manager choosing between two testing
approaches $A$ and $B$.
Each approach leads to a series of outcomes with different probabilities.
The probabilities come from a statistical analysis of the available data from using $A$ and $B$
in the past.
For instance, we may learn that approach $A$:
\begin{enumerate*}
    \item \label{aA:1} fails to produce bug-revealing tests in 3\% of its applications;
    \item \label{aA:2} generates too many redundant failing tests in another 3\% of applications;
    \item \label{aA:3} and works fairly successfully in the remaining applications.
\end{enumerate*}
According to CPT, each outcome's probability is weighted so as to reflect loss aversion or
other subjective phenomena.
For instance, if the losses associated with outcomes 1 and 2 above are very large,
the overall expected value of testing approach $A$ should be small because the person
in charge of the decision is more likely to avoid any risk of large losses.

An additional advantage of using CPT on top of a pure probabilistic model is that
the notion of subjective expected value of a decision is intuitive and understandable
by decision makers---building a kind of abstraction layer on top of 
a harder-to-interpret probabilistic model.

\subsection{Practical Significance in ESE}
Practical significance is important in an engineering discipline where research should inform ``solutions to practical problems''~\cite{shaw90}.\footnote{Still, one's definition of ``practical'' may differ from another's.} 
Indeed, the emergence of empirical software engineering 
was driven in part by the desire to identify effectiveness and significance in practice.
For example, one of the early guides to ESE research~\cite{kitchenham2002preliminary} 
highlights the need to ``differentiate between practical and statistical significance''.

Despite this early agenda, practical significance is rarely explicitly discussed in software engineering research publications~\cite{deOliveiraNeto2019Statistics}. 
When it is, there is a tendency to conflate it with statistical measures of strength of evidence.
Unfortunately, even when rigorous statistics are applied to carefully controlled experiments,
they may not generalize to realistic conditions~\cite{Stol2018,runkel1972,williams2019methodology}.

Other kinds of evidence (besides controlled experiments) 
may also support claims of practical significance.
For example, case studies
support generalization by providing instances that probe the boundaries
of the applicability domain~\cite{Yin.:2002}.
Another approach is framing software engineering as a design science~\cite{Wieringa2014,Storey2017_esem},
which focuses on the feedback loop between problem space and solution space. 
Then, practical significance can be tested when a solution is deployed in a practical setting. 

Another important aspect of generalizability, and hence practical significance, that 
can be evaluated empirically is \emph{scalability} to realistic settings~\cite{Wieringa2014}. 
Even more fundamental is that the investigated \emph{problem} must be important to industry practitioners. 
This is tricky because the practitioner's view is typically tied to a specific problem context.
For example, \cite{Tantithamthavorn2018icse-seip} 
mentions the example of a defect prediction study that focuses on predicting defects. 
Most of the value for practitioners does not lie in defect prediction \textit{per se} but
rather in how this information can guide decisions and explain the origin of defects. 
In this paper, we use CPT to model such connections between a technique's raw performance
and its value in terms of decision making.

Effect sizes and replications are  the main tools of statistical analysis that can support assessing practical significance. 
Effect size measures connect the outcomes of statistical tests to a measure of real-world impact. 
There are numerous effect size approaches, including $R^2$ in regression models (portion of variance explained), regression coefficients, Cohen's $d$ (standardized mean difference), Hedge's $g$, Cliff's $\delta$ (differences for ordinal data), and odds-ratios to capture relative effects.\footnote{See https://rpsychologist.com/cohend/ for an interesting visual tool for exploring effect sizes using Cohen's $d$.}

When researchers translate a raw effect size number into practical terms (moving from the estimate to population effects), they often use Cohen's t-shirt sizing approach (S,M,L). 
Correll \cite{Correll2020} describes in detail the problems with this categorical approach to effect size, including inconsistent bin thresholds, and Cohen himself said binning was a last resort:  
 ``\emph{contextual}, subjective judgment of observed effect sizes must be made and a ritualized interpretation avoided (emphasis ours, as quoted in~\cite{Kampenes2007})".  
The effect size is important, because frequentist analysis uses estimates of expected effect size to determine the appropriate sample size, given a particular power threshold (typically 80\%). 
Notwithstanding these existing and well-investigated effect size metrics, in empirical software engineering research reporting effect sizes is not widespread~\cite{Kampenes2007}, albeit, as can be seen from \cite{deOliveiraNeto2019Statistics}, a positive trend is visible (when not controlling for the possibility that journals publish more papers on a yearly basis). 
More importantly, effect size ignores the context of decision making. 
A raw number reflecting (for example) the standardized difference of means is hard for practitioners to interpret and must be contextualized. Decision analysis is the use of results from statistical inference to support decision-making in a specific context, for which effect size or, as in this paper, posterior predictive distributions, serve as inputs.


\section{Connecting Statistical Significance with Practical Significance}
\label{sec:experiment_bda}

Statistical significance in itself does not imply that an effect has practical value or utility.
Rather, practical significance is assessed on top of a statistical model that summarizes the data and its variability.
As a concrete case, to show how such an analysis can be performed, we first analyze a previously published study
on testing practices~\cite{AfzalGITAB2015et}, following 
the first four steps in Fig.~\ref{fig:our_approach}. Below, we provide further details on this case study, its statistical modeling and significance, as well as noting differences between different statistical approaches.

\subsection{The Case Study}
The study described in~\cite{AfzalGITAB2015et}
compares exploratory testing 
and testing based on documented test cases. 
In exploratory testing, developers are 
free to experiment with the system in an interactive fashion.  
In test-case based testing, developers
are required to document their work by writing test cases and oracles,
which can be re-executed at any later time. 

The experiment we consider here involved 35 developers,
classified into two categories according to their experience 
(years spent as software developers):
12 less experienced testers, and 23 more experienced ones. 
The developers included both industrial practitioners 
and software engineering students.
Unsurprisingly, on average, 
members of the former group had more experience than members of the latter;
but some students were still classified as ``more experienced'' 
and some practitioners as ``less experienced''.

Following a $2 \times 2$ cross-over experimental design,
each developer was randomly assigned to
one of the two testing approaches (exploratory or test-case based),
so that each combination of approach and experience
included a similar number of developers. After the first session the developers switched techniques.

During the experiment,
developers had to apply their assigned testing approach to
find as many faults as possible when testing an integrated development 
environment (\texttt{jEdit}). The number of faults found during the 
allotted time (two 90-minute sessions) measured each developer's \emph{effectiveness}---which
should reflect, thanks to randomization and experimental design, 
the intrinsic effectiveness of different testing techniques.
Table~\ref{tab:statistics_data} summarizes the experimental data.

\begin{table}
    \centering
    \caption{Summary statistics about the experimental data. 
    For each category of developers (low or high): 
    the number of developers, and the median, mean, standard deviation, minimum, and maximum number of faults each of them found during the experiments.}
    \label{tab:statistics_data}
    \begin{tabularx}{\columnwidth}{YcYYYYY}
        \toprule
        &   & \multicolumn{5}{c}{faults found through testing}\\
        \cmidrule{3-7}
        experience & $n$ & median & mean & sd & min & max\\
        \midrule
        low & 12 & 3 & 4.3 & 4.3 & 0 & 20 \\
        high & 23 & 5 & 6.0 & 5.3 & 0 & 18 \\
        \midrule
        any & 35 & 4 & 4.9 & 4.7 & 0 & 20 \\
        \bottomrule
    \end{tabularx}
\end{table}

\subsection{Statistical Modeling}
\label{sec:experiment_bda:design}

The main goal of our statistical analysis
is inferring a probability distribution of the number of
faults detected by developers testing the system.
Thus, the outcome is a natural number ($\mathbb{N}$) \nfaults,
which depends on two categorical predictors
capturing the testing \approach (exploratory or test-case based)
and the developer's \experience (low or high) in each trial.


\mypara{Population-level effects model}
Let us first consider a general linear model $\model_1$.
Since \nfaults is a non-negative integer variable representing a count, 
a Poisson distribution is a suitable \emph{likelihood} distribution~\cite{mcelreath15statrethink}
relating predictors and outcome, as shown in Eq.~\eqref{m1:like} in $\model_1$'s definition below. 
Both \nfaults and $\lambda$ have a subscript $i$,
which makes it explicit that we evaluate the model for each \emph{subject}
$i$ among all 35 developers (the dependence on the subject is usually left implicit;
we make it explicit so that it is clear what depends on the subject and what does not).
The Poisson distribution's rate $\lambda$
is the log-linear function (Eq.~\eqref{m1:link}) of predictors \approach
and \experience---each modeled as a binary indicator variable
for the two possible approaches and experience levels.
Finally, to apply Bayes' theorem we need to define \emph{priors}
for $\model_1$'s parameters $\alpha$, $\beta_a$, and $\beta_e$.
A standard choice, which works well in most cases,
is a weakly-informative prior such as a normal distribution
with zero mean and moderate standard deviation, as shown in Eq.~\eqref{m1:prior}.
This prior does not bias the effect that the predictors may have towards
positive or negative values,
and it still allows for a large range of possible parameter values---even
though extreme values (corresponding to very large effects) are increasingly unlikely.\footnote{Choosing even weaker priors, such as completely flat ones, would not affect the overall inference but may make sampling less efficient.}
Here is the overall definition of $\model_1$.

\begin{IEEEeqnarray}{rCl}
  \mr{faults}_i &\sim& \mr{Poisson}(\lambda_i)\label{m1:like}\\
  \log(\lambda_i) &=& \alpha + \beta_a \cdot \mr{approach}_i + \beta_e \cdot \mr{experience}_i\label{m1:link}\\
  \alpha,\beta_a,\beta_e &\sim& \mr{Normal}(0,1.5)\label{m1:prior}
\end{IEEEeqnarray}

Fitting $\model_1$ using the data gives 
a joint probability distribution on the parameters
$\alpha, \beta_a, \beta_e$, which
together identify a \emph{posterior} probability distribution
of \nfaults given the data.
According to Eq.~\eqref{m1:link}, 
the parameters connecting predictors to outcome
are the same for every subject in the experiment 
(that is, they do not depend on $i$);
thus $\model_1$ is a \emph{population-level effects} model.\footnote{Population-level effects are often known as ``fixed'' effects.}

\mypara{Varying effects model}
Bayesian analysis stresses the importance of modeling
data under different assumptions.
Therefore, let us consider ways of extending $\model_1$
into a finer-grained $\model_2$, which may capture
additional characteristics of the data under analysis.

By looking into the data more closely, we note that over $18\%$ of the developers found no faults during the experiments; 
that is, outcome $\nfaults = 0$
occurs more frequently than what a Poisson distribution predicts.
To account for this, we use a \emph{zero-inflated} Poisson
distribution as likelihood in $\model_2$.
As shown in Eq.~\eqref{m2:like},
such a distribution depends on a rate $\lambda$, like in a regular Poisson,
but it may produce a count of zero with probability $p$ in each draw.
However, unlike $\lambda$, for parameter $p$
we use a logit function.
As shown in Eq.~\eqref{m2:link-p}, 
we assume that only variable \approach
may affect $p$ since all cases of developers finding zero faults
occurred when using test-case based testing.

The other modeling assumption of $\model_1$ that we want to reconsider
are the population-level effects.
To account for the possibility that each subject $i$
may have different intrinsic skills at finding faults,
we add an intercept term $\alpha_{\mr{SUBJECTS}_i}$
to our linear regression (Eq.~\eqref{m2:subj}).
This term
represents the ``baseline'' contribution of each subject 
to the number of faults that are detected.
In summary, this makes $\model_2$ a \emph{varying effects} model.\footnote{Varying effects are often known as ``random'' effects.}

As in $\model_1$, $\model_2$'s priors are weakly informative;
the standard deviation $\sigma_s$, for the subject-specific intercept 
$\alpha_{\mr{SUBJECTS}_i}$, follows a half Cauchy distribution
(Eq.~\eqref{m2:prior-subjects}),
which is a common choice~\cite{Gelman2006priors} for a standard deviation (a non-negative real value).
Here is the overall definition of $\model_2$:

\begin{IEEEeqnarray}{rCl}
  \mr{faults}_i &\sim& \mr{ZIPoisson}(p_i, \lambda_i)\label{m2:like}\\
  \log(\lambda_i) &=& \alpha + \beta_a \cdot \mr{approach}_i + \beta_e \cdot \mr{experience}_i\label{m2:link}\\
  &+& {} \alpha_{\mr{SUBJECTS}_i}\label{m2:subj}\\
  \log(p_i) &=& \alpha_p + \beta_p \cdot \mr{approach}_i\label{m2:link-p}\\
  \alpha_{\mr{SUBJECTS}} &\sim& \mr{Normal}(\mu_s,\sigma_s)\label{m2:prior-subjects}\\
  \alpha,\beta_a,\beta_e &\sim& \mr{Normal}(0,1.5)\\
  \alpha_p,\beta_p,\mu_s &\sim& \mr{Normal}(0,1.5)\\
  \sigma_s &\sim& \mr{Cauchy}^+(0,1)
\end{IEEEeqnarray}


As for $\model_1$, fitting $\model_2$ using the data gives 
a joint probability distribution on the parameters
$\alpha, \beta_a, \beta_e, \alpha_{\mr{SUBJECTS}}, \alpha_p, \beta_p$, 
which together identify a \emph{posterior} probability distribution
of \nfaults given the data.

More precisely, 
the posterior is derived using 
statistical frameworks such as Stan~\cite{stan},
which work by sampling numerical approximations.
Therefore, in practice, the posterior is not an analytical expression but
rather a computational object.

\mypara{Model comparison}
We introduced $\model_2$ as a refinement of $\model_1$
based on some features of the data under analysis.
This should make $\model_2$ fit the data better, but
it may also increase the risk of \emph{overfitting}.
Fortunately, Bayesian analysis offers techniques
to quantitatively compare models 
selecting those that achieve the ``best'' trade-off
between fitting the data accurately while avoiding overfitting.\footnote{Posterior predictive checks were conducted for each model to judge the degree of fit.}
To this end, we use an {information criterion}.
An information criterion is a relative measure of how well
a model performs out-of-sample predictions compared to other competing models.

In our case, the PSIS-LOO state-of-the-art information criterion~\cite{loo}
indicates that $\model_2$ outperforms $\model_1$ in out-of-sample
prediction 
(see the replication package for a thorough explanation of this part of the analysis).
Therefore, we use $\model_2$ in the rest of our analysis, 
since it captures trends in the data better, while avoiding overfitting.

\subsection{Statistical Significance}
The posterior is a probability distribution over the 
parameter space,
which quantifies the degree of belief in each possible
combination of parameter values.
Therefore, it captures probabilistic features of the process we are analyzing,
namely how fault detection is affected by the chosen testing approach
and the developer's experience.

\begin{figure}
    \centering
    \includegraphics[width=\columnwidth]{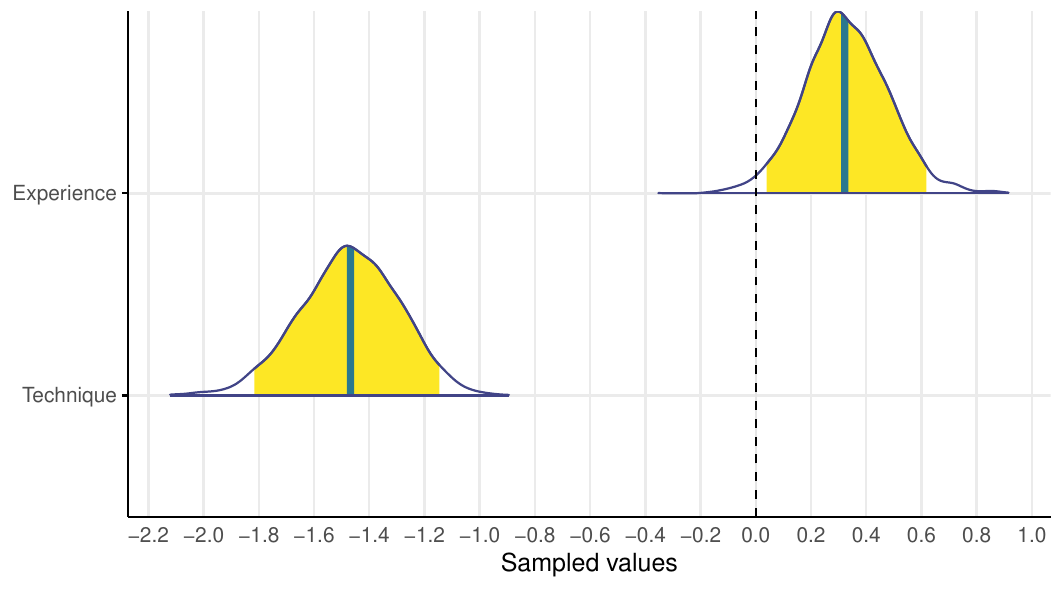}
    \caption{Posterior marginal probability distributions of $\beta_e$ (\experience, top) and $\beta_a$ (\approach, bottom named `Technique'). The thick lines mark the
    medians, and the yellow areas cover 94\% of probability.}
    \label{fig:exp_post_density}
\end{figure}

From the posterior's joint probability distribution
we can compute the \emph{marginals} for the parameters of interest.
For example, Fig.~\ref{fig:exp_post_density}
displays the marginals of coefficients $\beta_a$ and $\beta_e$.
Parameter $\beta_a$ models the effects of the testing \approach
used by each developer (named `Technique' in the figure)\footnote{Note that we use 94\% rather than 95\% intervals in the figure since the former is customary in the field of CPT.}.
Since $\beta_a$ is estimated to be very clearly negative, it
means that the chosen testing approach consistently correlates with 
the number of faults that are found.
Since \approach is a binary variable, with $0$ corresponding to exploratory testing and $1$ corresponding to test-case based testing,
a negative coefficient $\beta_a$ means that exploratory testing
is associated with \emph{more} faults being detected.
In contrast, parameter $\beta_e$ is likely positive.
Since it models the effects of developer \experience (another
binary variable with $0$ corresponding to low experience)
it means that more experienced developers tend to find more faults.
However, $\beta_e$'s distribution in Fig.~\ref{fig:exp_post_density}
has a non-negligible overlap with zero.
Hence, we have weaker confidence in the significance 
of experience than we have in the significance of the testing approach.

\begin{table}
    \centering
    \caption{Summary statistics of all population-level parameters in $\model_2$:
    mean (estimate), standard deviation (error), and lower and upper endpoint
    of the 94\% probability interval (credibility interval)
    of the parameter's posterior.}
    \label{tab:exp_sampled_posterior}
    \begin{tabular}{crrrr}
        \toprule
        \multicolumn{1}{c}{parameter} & \multicolumn{1}{c}{mean} & \multicolumn{1}{c}{std.\ dev.} & \multicolumn{2}{c}{94\% CI} \\
        \midrule
        $\alpha$  &  1.95 &  0.10 & 1.75,& 2.13 \\
        $\beta_a$ & -1.47 &  0.18 & -1.83, & -1.13 \\
        $\beta_e$ &  0.33 &  0.15 & 0.03, & 0.63 \\
        $\alpha_p$& -4.61 &  1.35 & -7.75, & -2.56 \\
        $\beta_p$ &  3.39 &  1.61 & 0.56, & 6.80 \\
        $\sigma_s$&  0.29 &  0.09 & 0.10, & 0.45 \\
        \bottomrule
    \end{tabular}
\end{table}

Table~\ref{tab:exp_sampled_posterior}
summarizes the posterior of all population-level parameters
of $\model_2$.
Through them, we can analyze other features of our fitted model.
For instance, $\beta_p$ is clearly positive,
which means that test-case based testing is
associated with a higher probability of detecting no faults.

\mypara{Towards practical significance}
By analyzing a posterior probability distribution 
we can move from statistical significance to 
practical significance.
Since a full distribution is available, 
we are not limited to measuring probability intervals of the model's parameters,
but we can also calculate probabilities of \emph{outcomes}---that is
what is the expected number of \emph{detected faults}
in different scenarios.

\begin{figure}
    \centering
    \includegraphics[width=\columnwidth]{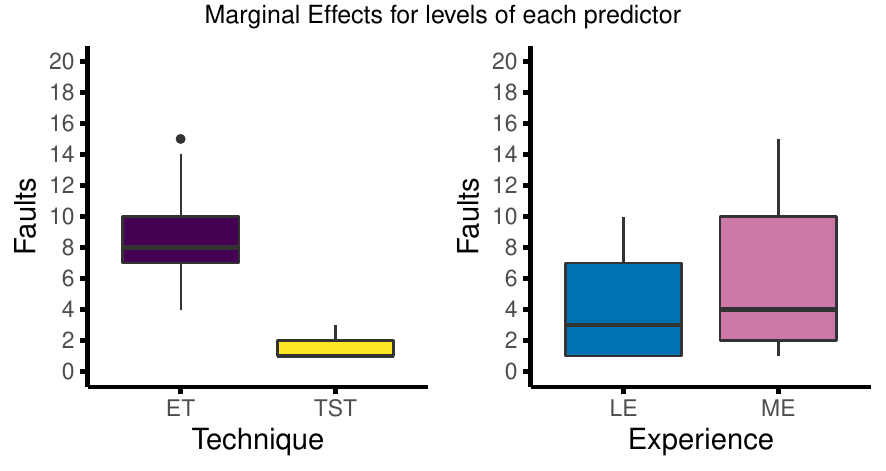}
    \caption{Left: expected number of detected faults with 94\% probability
    for developers using exploratory testing and test-case based testing.
    Right: expected number of detected faults with 94\% probability
    for developers with low experience and high experience.}
    \label{fig:exp_marg_plot}
\end{figure}

Concretely, we derive different marginal distributions from the posterior 
according to specific usage scenarios that we may want to analyze.
For example, to estimate the expected number
of \nfaults that would be detected by a developer
using exploratory testing ($\approach = 0$)
or by one using test-case based testing ($\approach = 1$).
Figure~\ref{fig:exp_marg_plot} (left figure) shows the ranges spanning a 94\% 
probability interval, which not only confirm that
exploratory testing is expected to be more effective but 
quantify the expected difference (roughly three times as many faults found).

A similar analysis comparing developers with different experience,
in Fig.~\ref{fig:exp_marg_plot} (right figure),
suggests instead that the fault-detection performance
of developers with different experience can still be very similar as
the two intervals have a large overlapping---even though
more experienced developers are slightly more effective on average.

Table~\ref{tab:prediction} reports the
same information numerically, and extends the analysis to other scenarios
such as comparing the effects of developer experience on the number of faults detected using exploratory testing.
In Sect.~\ref{sec:experiment_cpt} we will show these probabilities
buttress a rigorous analysis of practical significance.


\begin{table}
    \centering
    \caption{Expected number of
    faults detected 
    for different combinations of predictors in $\model_2$.
    Each row reports the range of \nfaults corresponding to 94\%
    probability and the mean on the posterior.
    }
    \label{tab:experiment_predictions}
    \label{tab:prediction}
    \begin{tabular}{llrlr}
        \toprule
        \multicolumn{1}{c}{developers} & \multicolumn{1}{c}{fixed predictors} & \multicolumn{2}{c}{94\% CI} & \multicolumn{1}{c}{mean} \\
        \midrule
        low experience  & $\!\!\!\!\!\begin{array}{l}\experience = 0 \end{array}$ & 1,&  \ \,8 & 4.02 \\
        high experience & $\!\!\!\!\!\begin{array}{l}\experience = 1 \end{array}$  & 1, & 11 & 5.67 \\
        exploratory testing & $\!\!\!\!\!\begin{array}{l}\approach = 0\end{array}$ & 6, &11 & 8.27 \\
        test-case testing  & $\!\!\!\!\!\begin{array}{l}\approach = 1\end{array}$ & 1,  & \ \,2 & 1.42 \\
        exploratory and low & $\!\!\!\!\!\begin{array}{l}\approach = 0 \\ \experience = 0 \end{array}$ & 6,  & \ \,8 & 6.92 \\
        exploratory and high & $\!\!\!\!\!\begin{array}{l}\approach = 0 \\ \experience = 1 \end{array}$ & 8, & 12 & 9.61 \\
        \bottomrule
    \end{tabular}
\end{table}

\subsection{Bayesian vs. Frequentist Analysis of Significance}
\label{sec:bvf-significance}

Overall, the high-level results of our Bayesian reanalysis
are consistent with those of the original study~\cite{AfzalGITAB2015et}:
exploratory testing performs significantly better than
test-case based testing when looking at the number of faults found;
the impact of experience is 'clearly' significant in the original study, while in the reanalysis one can question that.
Unlike our reanalysis, the original study used frequentist statistics---similarly to
previous analyses of the same processes~\cite{ItkonenM14exp, Itkonen13ML}.

Even though the big picture does not change---and there is no reason
it should---the distinctive features of Bayesian statistics make
our reanalysis results 
more directly useful
to assess \emph{practical} significance in a robust and insightful way.
The Bayesian emphasis on modeling entails that we
could consider and compare different competing statistical models
on the grounds of their characteristics and performance.
More important, features of the chosen model $\model_2$
strengthen our understanding of the studied phenomenon:
zero-inflation indicates that a significant portion of the
trials found no faults;
a varying-effects term accounts for the significant individual
differences among developers.
These model features do not affect the conclusion that one
testing approach performs significantly better than the other;
but they help quantify the relative weight of different
factors more precisely and in terms of phenomena
in the actual problem domain.

The other key feature of Bayesian analysis
that strengthens our understanding of the studied phenomena
is the capability of deriving probability distributions of the outcome variables.
In our case, we derived the expected number of \nfaults a
pool of developers with certain characteristics would find.
This expresses the ``significance'' of a
certain difference between treatments in terms of
measures that are relevant in practice in the domain of testing processes.

As we show in the next section, this feature 
is also the basis to combine a statistical model of the data
with features of how humans assess probabilities
and make decisions---leading
to an all-round assessment of practical significance.

\section{Arguing Practical Significance}
\label{sec:experiment_cpt}

We want to frame practical significance in
a way that it supports \emph{choices} between alternatives.
Based on our case study, 
we imagine a manager who can select a testing approach
(exploratory or test-case based)
and developers with lower or higher experience.
To support the manager's choices,
we quantify the expected \emph{utility} of each choice: 
a monetary value that approximates the gains (if positive)
or losses (if negatives) that are likely to derive from that choice.

Even though our example can be seen as simplistic, it is grounded on a need that partners in industry had, when deciding between two test techniques~\cite{AfzalGITAB2015et}. However, more complex decisions can also be made, in particular when involving multiple stakeholders with conflicting views. But, in those cases one would need to look at game theory as the underlying decision-making framework~\cite{nash}. In our case, one could rather see an alternative, and if you will, more straightforward, approach. 

\subsection{Value and Weight Functions}
A simple approach would just use the probabilities of 
different outcomes (computed from $\model_2$)
to average the value $\nu(x)$ of each possible outcome $x$ of decision $X$.
Instead, we use cumulative prospect theory (CPT)
to ``adjust'' the probabilities so that they reflect
how humans are likely to perceive alternatives.
The expected ``subjective'' utility $\expect_U(X)$ of decision $X$ 
is given by the weighted average,\footnote{If the outcomes are a continuum, the average should be computed using an integral and cumulative probabilities.}

\begin{equation}
    \expect_U(X) \ =\ 
    \sum_{x \in X} w(P(x)) \cdot \nu(x).
    \label{eq:utility}
\end{equation}
where $w$ is a \emph{weight function} that adjusts the
probability $P(x)$ of each outcome.

In Eq.~\eqref{eq:utility} above,
$P$ can be computed from our posterior probability distribution.
As weight function we can use a standard function proposed by Tversky and Kahneman~\cite{Tversky1992cpt}. The idea of this function is that intermediate probabilities 
are flattened as if they were similar,
whereas probabilities closer to the extremes are accentuated.
This reflects the human perception that tends to conflate small probabilities with
impossibility, and large probabilities with certainty.

We are left with providing a definition of $\nu(x)$
for each possible outcome $x$.
In our case study, outcome $x$ corresponds to $\nfaults = x$,
that is $x$ faults detected during a testing session.
We can associate a monetary value to each outcome
based on the costs and benefits that come with it:
\begin{equation}
\label{eq:experiment_tpcost}
    \nu(\nfaults = x)\ =\ (S \cdot x) - (C \cdot h)
\end{equation}
where $S$ are the savings for each fault found,
$C$ is the hourly pay of a developer,
and $h$ is the number of hours of a testing session.
Since our study's experiment involved $2 \times 90$ minute sessions,
we set $h = 3$.
The manager of a company could suggest values for $S$ and $C$.
In this section, we take $S = \$150$, $C^- = \$100$ for low-experience
developers, and $C^+ = \$200$ for high-experience developers.
These values are realistic values for a small software company in Sweden.

\subsection{Utility of Different Choices}

Consider three possible choices a manager
has to make, and compute the expected utility according to Eq.~\eqref{eq:utility}.

\begin{description}
\item[approach:] the manager chooses whether developers 
use exploratory testing or test-case based testing;
the available developers are a mix of low-experience and high-experience
\item[experience:] the manager chooses whether to hire
low-experience or high-experience developers to do testing;
they will use a mix of exploratory and test-case based testing
\item[exploratory:] the manager chooses whether to hire
low-experience or high-experience developers to do exploratory testing
\end{description}

Whenever there is a mix of options, we assume it
reflects averages in our data (i.e., the sample is representative of the population).

\subsubsection{Choosing the Approach}
In this scenario, \textbf{approach} is
the choice is between using \emph{exploratory} testing or \emph{test-case} based testing.
We use $\overline{C} = \$134.38$
as hourly developer cost.
This is the average per-person cost in a pool
of 35 developers---23 with low experience and 12 with high experience
as in the dataset.

According to Eq.~\eqref{eq:utility},
the expected utility of choosing exploratory testing is $\expect_U(\textit{exploratory}) = \$454.3$;
the expected utility of choosing test-case based testing is
$\expect_U(\textit{test-case}) = \$8.0$.
Hence, it is clear exploratory testing is likely to 
bring a much higher value.

\begin{figure}
    \centering
    \includegraphics[width = 0.45\columnwidth]{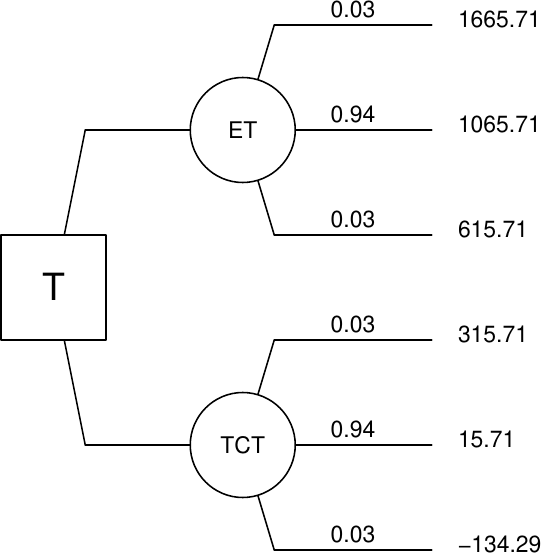}
    \caption{Utility for different choices in scenario \textbf{approach}: 
    the manager chooses whether developers 
use exploratory testing (ET) or test-case based testing (TCT).
    }
    \label{fig:experiment_choices_tech}
\end{figure}

In addition to expected utility, we 
can compute the utility value associated with outcomes 
with a certain probability.
As is customary in CPT, we split the probability unit interval
into three parts $3\%, 94\%, 3\%$
and compute utility for each sub-interval.\footnote{We like to present the tails of a distribution, and the tails should be much smaller than the bulk of the distribution.}
Figure~\ref{fig:experiment_choices_tech} 
shows the results for the scenario's \textbf{approach}:
using test-case based testing
would lead to significant gains ($\$315.71$) with 3\% probability;
to significant losses ($-\$134.29$) also with 3\% probability;
and to modest gains ($\$15.71$) in the vast majority of cases (94\% probability).
In contrast, using exploratory test-case based testing
makes any losses vanishingly unlikely to happen.

\subsubsection{Choosing the Experience}
\begin{figure}
\centering

\includegraphics[width = 0.45\columnwidth]{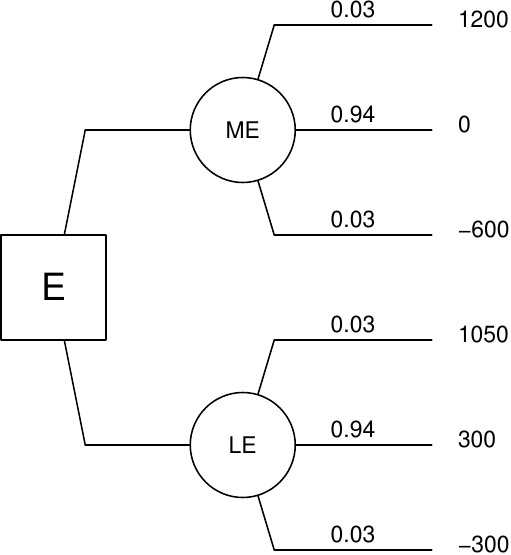}
\caption{Utility for different choices in the scenario for \textbf{experience}: 
    the manager chooses whether to hire developers 
with low (LE) or high (ME) experience.}
\label{fig:experiment_choices_exp}
\end{figure}

In this scenario, \textbf{experience} is
the choice between using developers with \emph{low} or \emph{high} experience.
According to Eq.~\eqref{eq:utility},
the expected utility of hiring low-experience developers 
is $\expect_U(\textit{low}) = \$304.50$,
and
the expected utility of hiring high-experience developers 
is $\expect_U(\textit{high}) = \$18$.
This means that, with the combination of exploratory 
and test-case based testing seen in the case study,
high-experience developers are not worth what they cost;
low-experience developers achieve a more favorable trade-off.
Figure~\ref{fig:experiment_choices_exp} shows the breakdown
of utility for different ranges of probability in the scenario regarding \textbf{experience}.

Remember that these results depend on the assumptions about the cost of undetected faults and the hourly cost of developers. If, for example, each detected fault
would bring a gain of $\$ 1000$---rather than $\$ 150$ as we have assumed so far---the expected utility of hiring low-experience developers would become
$\$6700$, smaller than the expected utility $\$9400$ of hiring high-experience developers.
In other words, if finding as many faults as possible is critical, 
even the modest performance advantage of high-experience developers
may be worth the higher costs.

\subsubsection{Choosing the Experience, Given Exploratory Testing}
\begin{figure}
\centering
\includegraphics[width=0.45\columnwidth]{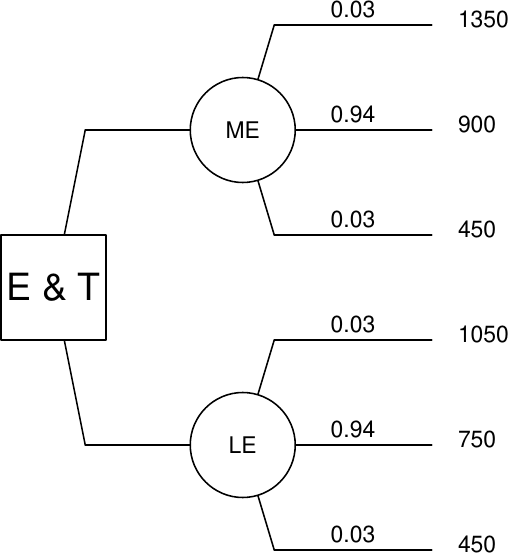}
\caption{Utility for different choices in the scenario \textbf{exploratory}: 
    the manager chooses whether to hire developers 
with low (LE) or high experience (ME) to perform exploratory testing.}
\label{fig:experiment_cp3}
\end{figure}

In this scenario, \textbf{exploratory} is
the choice, again, between using developers with \emph{low} or \emph{high} experience, but all of them use exploratory testing.
In this case,
the expected utility of hiring low-experience developers 
is $\expect_U(\textit{low + exploratory}) = \$750$;
the expected utility of hiring high-experience developers 
is $\expect_U(\textit{high + exploratory}) = \$900$.
More experienced developers are worth their higher pay in this scenario;
but the difference is small, and hence
a manager may also include low-experience developers
if high-experience ones turn out to be hard to find or unavailable.
Figure~\ref{fig:experiment_cp3} shows the breakdown
of utility for different ranges of probability in the scenario of \textbf{experience}.

\subsection{Usage of CPT to Foster Practical Significance}

In summary, we used CPT to evaluate the practical impact of our investigation with respect to both the testing approach and the experience of testers. We designed three choice problems related to the factors investigated in our experiment, and provided different outcomes for each choice problem, along with a suggested decision based on the utility of those choices.

Cumulative prospect theory allows us to explicitly discuss and present practical significance by creating choice problems based on Bayesian analysis. 
The original study did not include a discussion about the practical impact of its results in connection to the costs of a fault or other contextual data from industry. Using our approach
in a hypothetical context with costs for faults and salaries, 
we could instead 
illustrate how statistical predictions and CPT support decision making. 

We should be clear that our values for costs of faults and salaries 
are arbitrarily chosen---though based on estimates of the Swedish job market---and could change dramatically according to the software's domain,
the characteristics of the company, and many other factors~\cite{BoehmB01top-10}.
Our only assumption is that
it is somewhat possible to estimate such costs by collecting the necessary information
from research, practitioners, and common knowledge.\footnote{%
When a precise estimate is unavailable, 
the model could also incorporate \emph{uncertainty} in the estimates
as probability intervals, and calculate how the uncertainty in the estimates translates to uncertainty in the outcomes.
Interested readers can try this out in our analysis using the supplementary material.}
Whenever this assumption is satisfied, this approach is applicable.
\section{Validation with Practitioners}
\label{sec:validation}

Section~\ref{sec:experiment_cpt}
demonstrated our approach of applying cumulative prospect theory (CPT)
on a posterior probability distribution
to argue significance in terms of costs and benefits
in practice.
%
On the other hand, 
statistics alone 
can---and often is---used to discuss significance
from a statistical perspective.
To ascertain whether practitioners and domain experts
do indeed find a presentation of significance in terms
of utility clearer than a traditional statistical analysis,
we conducted a qualitative empirical validation.
The rest of this section describes its design
and results.

\subsection{Validation Design}
The overall goal of the validation is to compare
two different ways of framing significance results:
\begin{enumerate*}
    \item using standard (frequentist) statistical techniques; and
    \item using our combination of Bayesian analysis and CPT.
\end{enumerate*}
The comparison is from the point of view of stakeholders
using statistically significant results to make a decision.

\mypara{Participants}
We contacted 22 managers working at two large companies in Sweden
who agreed to participate in this validation.
The participants are a convenience sampling, and
come from both middle (15 participants) and upper (7 participants) 
management positions.
None of the participants were involved in this research,
nor in the original analysis of testing approaches~\cite{AfzalGITAB2015et}.

\mypara{Instruments}
We provided all participants
with two summaries of the comparison between
exploratory and test-case based testing:
\begin{enumerate*}
    \item a summary based on the statistics and results of 
    the original study~\cite{AfzalGITAB2015et} (using frequentist statistics);
    \item a summary based on the statistics and results of the present paper's re-analysis (using Bayesian statistics and CPT).
\end{enumerate*}
Participants were asked to read the summaries and answer two yes/no 
questions about each of the summaries:
\begin{description}
    \item[approach:] Based on the presented information, would you use exploratory testing?
    \item[experience:] Based on the presented information, would you use more experienced testers?
\end{description}

After answering each question with yes or no,
participants also rated, on a 1--5 Likert scale, 
how confident they were in their answer 
(1: not confident at all; 5: completely confident), i.e., the construct used to assess which approach was better.

The questionnaire material was prepared by two of the authors
and validated by two other authors.\footnote{The questionnaire is available in the replication package.}
The hourly costs for junior\slash senior developers and the costs for a bug was set according to the participating companies' input, i.e., \$$60$, \$$70$, and \$$10,000$, respectively.
The questionnaire was then used at two workshops.
One of the authors was available to answer any questions about the questions or the summaries.

We only retained 18 questionnaires out of 22---those
that had all required information filled in.



\begin{figure}
    \centering
    \includegraphics[width=\columnwidth]{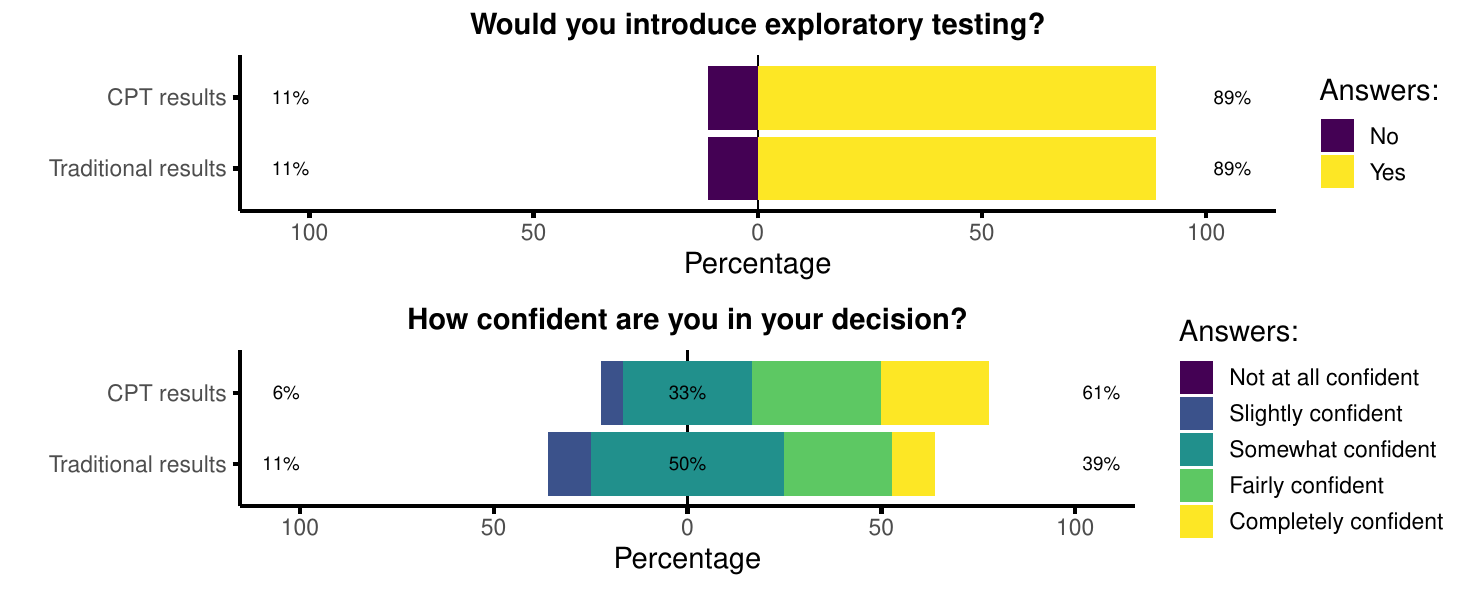}
    \caption{Responses from 18 subjects about question \textbf{approach}:
    ``Would you use exploratory testing?''}
\label{fig:validation_tech}
\end{figure}
\begin{figure}
    \centering
    \includegraphics[width=\columnwidth]{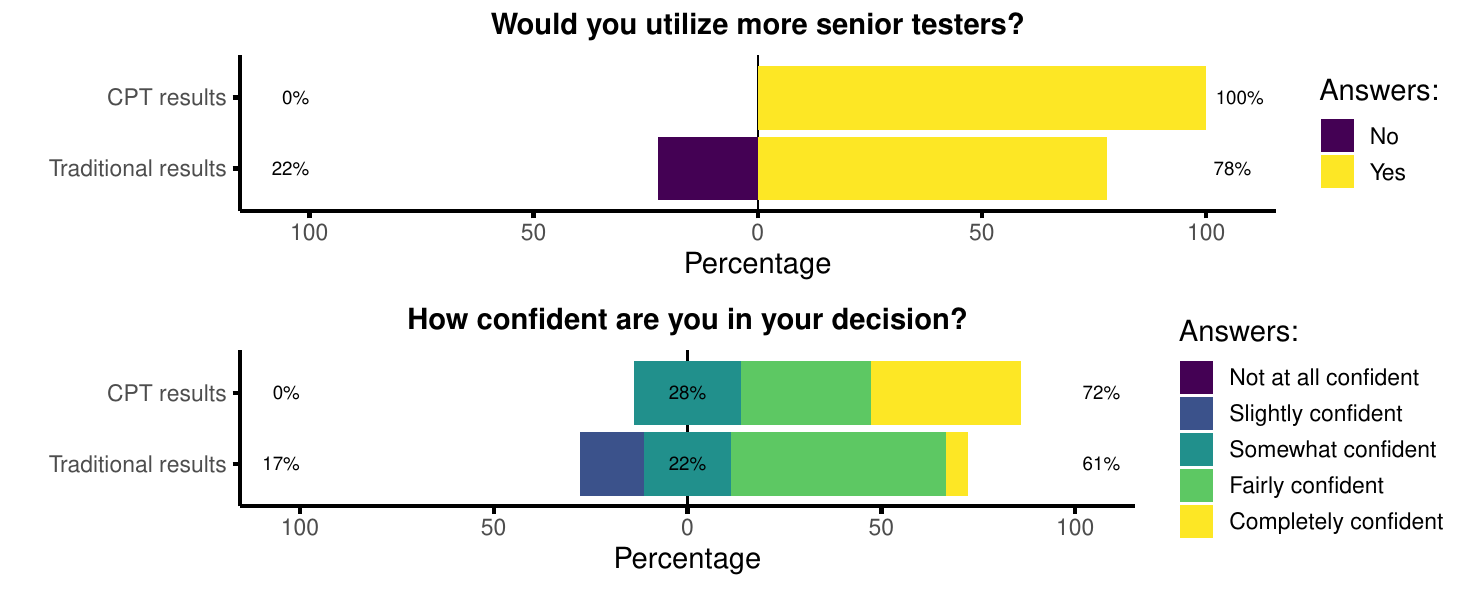}
    \caption{Responses from 18 subjects about question \textbf{experience}:
    ``Would you use more experienced testers?''}
    \label{fig:validation_exp}
\end{figure}

\subsection{Validation Results}
The results of the questionnaires are summarized using diverging bar charts in Figs.~\ref{fig:validation_tech}--\ref{fig:validation_exp}.
Each horizontal bar is centered on the middle point of the Likert scale (value 3)
and spans the percentage of participants answering 
with low (values 1 and 2 on the Likert scale spanning the left of the middle point)
or high (values 4 and 5 on the Likert scale spanning the right of the middle point).

Figure~\ref{fig:validation_tech}
summarizes the results concerning \textbf{approach}.
In this case, the decisions made by participants do not depend
on how the data is presented.
However, the summary based on our approach combining
Bayesian models and CPT tends to increase the confidence
in the decision:
61\% of participants were fairly or completely confident---instead of
39\% with the ``traditional'' summary.

Figure~\ref{fig:validation_exp}
summarizes the results concerning \textbf{experience}.
In this case, the summary based on our approach convinced
all participants to choose more experienced testers,
whereas the ``traditional'' statistical summary convinced only 78\% of them.
As in answers to the other question,
Bayesian models and CPT tend to increase the confidence
in the decision:
39\% of participants were completely confident---instead of
6\% with the ``traditional'' summary.
In both cases, confidence in answering this question was 
higher than in answering question \textbf{approach}.

By analyzing the recorded discussions after the session we received a qualitative, richer, view on the perception the subjects had concerning the two approaches, which we define as Case $1$ (the original study's results) and Case $2$ (the results we propose in this paper).

First, it is worthwhile to note that the subjects were only asked, in a group session, two questions to start the discussions concerning Case $1$ and $2$: How do you make sense of the results in [Case $1$ or Case $2$]? Second, the subjects spent approximately twice as much time discussing Case $2$. Quite simply, the subjects exhausted Case $1$ much sooner. If we consider two very representatives quotes concerning Case $1$, we will see why this could be the case:

\begin{quote}
    Quite difficult to understand. In particular the difference between more or less experienced developers was very unclear in [Case $1$].  
\end{quote}

and,

\begin{quote}
    I was more looking into the summary [the conclusion] \ldots it said it was no significant difference [between more or less experienced developers].
\end{quote}

The first quote, which is representative of many comments, indicates that Case $1$ is more difficult to understand. The second quote could be an indication that Case $1$'s use of frequentist statistics, where we fall back on the arbitrary 95\% significance level, did not encourage them to look further.

Concerning Case $2$ the following quotes paint another picture:

\begin{quote}
    Very clear. It becomes so easy [\ldots] makes me instead think about other things [\ldots] how was data collected?
\end{quote}
    ,
\begin{quote}
    The example [Case $2$] made me start thinking more about other things that could affect the effectiveness [of the two techniques].
\end{quote}
    , and
\begin{quote}
    We can look at a research paper and interpret it from our [the company's] perspective.
\end{quote}

This validation
provides some preliminary evidence that
presenting significance results using
a combination of Bayesian statistics and cumulative prospect theory
might help increase the confidence
in a decision between alternatives---even when
the outcome of a decision is not greatly affected
since the underlying data remains the same.


\section{Discussion}
\label{sec:discussion}

We have presented a method for how to assess and argue for the practical significance of empirical software engineering results. By combining Bayesian Data Analysis (BDA) with systematic evaluation of outcomes, using Cumulative Prospect Theory (CPT), different scenarios can be simulated and compared. While the posterior probability distribution from BDA summarizes the (scientific) knowledge gained by the research, the scenario simulation can help practitioners connect to concrete situations and, thus, increases the practical significance of the research while also informing decisions. This is supported by our evaluation with managers at two companies in the software industry. 

Applying this approach in other settings is straightforward. 
It applies anywhere there is a quantifiable outcome for a technique, e.g., effort estimation or bug detection. 
The caveat is that measurements are only as good as their construct validity and the precision of the measurement. 
In situations where it is very hard to quantify the value of different outcomes it will be hard to apply the technique.

To apply the approach, we follow the steps outlined in Fig.~\ref{fig:our_approach}. To choose value and weighting functions, an initial approach is to choose a function that is a simple linear\slash exponential\slash sigmoid, as these three categories capture most of the valuation functions we have seen. Asking practitioners for three or four values and selecting between these three categories should then allow a suitable value function to be fitted. Other approaches are beyond the scope of this article but could include fuzzy logic, for example \cite{Bykzkan2004}.

In situations where the data and\slash or model is more complex than the case we re-analysed here, e.g. when there are more factors or they interact in affecting the dependent variable(s), we argue that the proposed methodology should be relatively more valuable. This is because humans would have a relatively harder time judging the scenarios or understanding the effects if there are complex or non-linear interactions. Simulation and concrete metrics as a basis for comparison thus will be more important.

There is a risk that the simplicity encouraged by the use of CPT or, really, any method that considers few factors and assigns them simple numerical values, would lead software engineers and managers to not consider the many factors that are critical to real-world decision making~\cite{nwogugu2006further}. For an obvious example, in the case studied here, it would not be wise for a manager to simply prefer a more experienced engineer over a less experienced, without considering how they would fit into the development team. However, we argue that practitioners understand this and do not expect research to fully cater to their everyday situation. Also, we argue that this is a risk with the underlying study itself regardless of the way the collected data is then analyzed. However, we still advise caution in the claims that one makes based on the type of analyses proposed here; the additional clarity offered by summarizing specific outcomes in simple numerical values should not be misused.

In closing, we note that there is research on how to present probabilities and research results for better effect~\cite{wijeysundera2009bayesian} and that there has also been several criticisms of CPT~\cite{gigerenzer1991make,nwogugu2006further}. We leave it to future work to explore alternatives to CPT, in this regard.

\section{Threats to Validity}
\label{sec:threats}

The canonical structure 
used to discuss 
threats to validity
mainly targets experimental design and sub-sequence statistical analysis;
hence it does not fit this work very well.
Instead, we present our analysis of threats to validity
as ``bad smells'' of analysis work\cite{krishnaMMS2018smells}.

\begin{enumerate}
\item \textbf{Not interesting.} (Research that has negligible software engineering impact.) The problem of analyzing practical significance is
itself relevant and significant in practice,
as demonstrated by previous work on statistical analysis that we summarize in 
Sect.~\ref{sec:background}.

\item \textbf{Not using related work.} (Unawareness of related work concerning RQs and SOA.) Section~\ref{sec:background} also discusses the previous approaches to arguing practical significance, and their limitations.

\item \textbf{Using deprecated and suspect data.} (Using data out of convenience.) 
Our case study is a reanalysis of existing data,
which was recently analyzed in a peer-reviewed publication~\cite{AfzalGITAB2015et}.

\item \textbf{Inadequate reporting.} (Partial reporting, e.g., only means.) 
One of the key features
of Bayesian statistics, which we propose as a basis for a more thorough analysis of
practical significance,
is its focus on modeling complete probability distributions
instead of only point estimates.

\item \textbf{Under-powered experiments.} (Small effect sizes and little theory.) 
While we did not perform power analysis explicitly,
model comparison and other diagnostic techniques
of Bayesian data analysis are useful
to choose models based on their effectiveness
in practice and in theory.

\item \textbf{$\boldsymbol{p< 0.05}$ and all that.} (Abuse of null hypothesis testing.) Null-hypothesis testing is manifestly in contrast to the 
statistical modeling approach we propose.

\item \textbf{Assumptions of normality and equal variances.} (Impact of outliers and heteroscedasticity.) 
In our case study, we use a (multi-level) generalized linear model with a Poisson likelihood. Additionally, we employed multi-level modelling, which helps avoid overfitting and handle outliers appropriately.

\item \textbf{Not exploring stability.} 
We did not discuss them in the paper for brevity,
but we ran all recommended diagnostics 
of Bayesian analysis to
ensure that there are no stability problems in the
fitted models used in our study, this includes prior predictive checks.

\item \textbf{No data visualization.} 
We used data visualization to complement numeric data---focusing
on the key measures of interest.

\item \textbf{Not tuning}, \textbf{not exploring simplicity}, 
\textbf{and} \textbf{not justifying choice of learner (overfitting).}
In our case study, we explored two competing models,
and chose one based on both theoretical considerations
and the information criterion concerning out-of-sample predictive performance.
Thus, when tuning our models, we considered models of different complexity,
and we employed techniques to reduce the chance of overfitting. 
\end{enumerate}
\section{Conclusions}
\label{sec:conclusion}
We presented an approach 
to argue practical significance in empirical software engineering.
Our approach develops a Bayesian model of the data,
then it applies cumulative prospect theory on top of the model
to incorporate a quantitative notion of utility
and how probabilities are subjectively perceived by humans facing a decision.
We demonstrated the approach on data from
a previously published study comparing
exploratory testing to test-case based testing~\cite{AfzalGITAB2015et}.

To ascertain whether our presentation of statistically significant results is 
indeed accessible to practitioners,
we conducted a small-scale empirical validation where we asked managers 
to report their confidence in decisions concerning the study's results.
More precisely, we compared decisions informed by our presentation of practical significance to decisions based on the original frequentist statistical analysis~\cite{AfzalGITAB2015et}.
Our combination of Bayesian statistics and
cumulative prospect theory tends to increase
the decision makers' confidence.

In future work, 
we plan to demonstrate our approach on larger case studies,
and to perform more extensive validations of its usefulness in practice.



\section*{Replication Package}
\label{appA}



A replication package---including all data analyzed in the paper,
scripts used to perform the analysis,
and a Docker image with the tools to run the scripts---is available at:

\url{https://github.com/torkar/docker-b3}


\ifCLASSOPTIONcompsoc
  \section*{Acknowledgments}
\else
  \section*{Acknowledgment}
\fi

We express our thanks to Paul-Christian B\"urkner, Aki Vehtari, and Jonah Gabry, for the discussion on priors, and pointing us to literature discussing the choice of priors~\cite{SimpsonRMRS14}.


\IEEEtriggeratref{32}



\bibliographystyle{IEEEtran}
\bibliography{refs.bib}
\end{document}